\documentclass[preprint2, twocolappendix]{aastex631}

\usepackage{amsmath,amsmath,ulem}
\usepackage{tikz}
\usepackage{ifthen}
\usepackage[flushleft]{threeparttable}
\usepgflibrary{fpu}
\graphicspath{{./}{figures/}}

\tabletypesize{\footnotesize}
\begin{document}

\title{The Trojan-like Colors of Low-Perihelion Kuiper Belt Objects}
\author[0000-0003-4778-6170]{Matthew Belyakov}
\affiliation{Division of Geological and Planetary Sciences, California Institute of Technology, Pasadena, CA 91125, USA}
\correspondingauthor{Matthew Belyakov}
\email{mattbel@caltech.edu}
\author[0000-0002-8255-0545]{Michael E. Brown}
\affiliation{Division of Geological and Planetary Sciences, California Institute of Technology, Pasadena, CA 91125, USA}

\author[0009-0001-9229-4581]{Alya al-Kibbi}
\affiliation{Department of the Geophysical Sciences, University of Chicago, Chicago, IL 60637, USA}

\begin{abstract}
An important testable prediction of dynamical instability models for the early evolution of the Solar System is that Jupiter Trojans share a source population with the Kuiper belt. Concrete evidence of this prediction remains elusive, as Kuiper belt objects (KBOs) and Jupiter Trojans appear to have different surface compositions. We address the long-standing question of Trojan origin by finding a dynamical sub-population in the Kuiper belt with Trojan-like colors. Combining existing photometric data with our own surveys on Keck I and Palomar P200, we find that the low-perihelion ($q<30 $AU, $a>30 $AU) component of the Kuiper belt has colors that bifurcate similarly to the Jupiter Trojans, unlike Centaurs ($a<30 $AU) which have redder, Kuiper belt-like colors. To connect the Jupiter Trojans to the Kuiper belt, we test whether the distinct Trojan-like colors of low-perihelion KBOs result from surface processing, or are sourced from a specific population in the Kuiper belt. By simulating the evolution of the Canada-France Ecliptic Plane Survey synthetic population of KBOs for four billion years, we find that differences in heating timescales cannot result in a significant depletion of Very Red low-perihelion KBOs as compared to the Centaurs. We find that the neutrally-colored scattered disk objects ($e>0.6$ KBOs) contribute more to the low-perihelion KBO population rather than Centaurs, resulting in their different colors. 
\end{abstract}

\section{Introduction}
The solar system's early large-scale dynamical instability described by the many iterations of the Nice Model has had great success in providing a single resolution to many phenomena in the solar system, including the excited eccentricities of the giant planets \citep{2005Natur.435..459T} and the structure of the Kuiper belt and asteroid belt \citep{2008Icar..196..258L, 2010AJ....140.1391M} among many other properties of the solar system \citep{2018ARA&A..56..137N}. Since the initial publication of the Nice model in 2005, the nature of its dynamical instability model has evolved significantly. The first three landmark Nice model papers relied on Jupiter and Saturn acquiring excited eccentricities through slowly migrating through their mutual 1:2 resonance \citep{2005Natur.435..459T}. Resonance crossing was abandoned to due its inconsistency with the structure of inner solar system, as during the instability, Saturn's secular resonances would sweep out material needed to form the inner planets. The current model for the early instability involves the ejection of a fifth (or potentially more) giant planet by Jupiter, which suddenly shifts Jupiter's orbit inwards, avoiding the issues of granular migration and resonance crossing \citep{2012AJ....144..117N, 2012ApJ...744L...3B, 2018ARA&A..56..137N}.

One of the most robust predictions present in all versions of the Nice Model is that the Jupiter Trojan asteroids at 5.2 AU share a source population with the more distant Kuiper belt \citep{2005Natur.435..459T, 2005Natur.435..462M, 2010CRPhy..11..651M, 2018ARA&A..56..137N}. In the current iteration of the Nice Model, the giant planets undergo a series of dynamical instabilities, caused by the ejection of one or more planetary cores formed somewhere between Jupiter and Neptune \citep{2008Icar..196..258L, 2012AJ....144..117N}. At the time of these instabilities, small bodies that are in the Trojan region of any of the giant planets would be thrown onto now chaotic orbits \citep{2005Natur.435..462M}. In this picture, the current Jupiter Trojans could only have been captured during Jupiter's final large-scale instability. Jupiter Trojans are thus simply the objects that happened to be at Jupiter's L4/L5 points during its final ``jump'', captured into librating orbits. Meanwhile, planetesimals originating in the primordial Kuiper belt disk would be scattering throughout the solar system, forming the dominant population of small bodies in the outer solar system. Consequently, the Nice model implies that progenitor objects for the Kuiper belt also became the Trojans we see today \citep{2005Natur.435..462M, 2013ApJ...768...45N}. This prediction gives a clear observational test for the Nice model: if Trojans and Kuiper belt objects (KBOs) indeed have a shared origin, these two populations and should have common physical properties \citep{2005Natur.435..462M,2009Icar..202..310M,2013ApJ...768...45N,2018AJ....155...56J}. However, observational backing for a shared origin remains elusive, leaving a fundamental gap in the evidence for the Nice model's predictions regarding the evolution of the outer solar system.

One piece of evidence that supports the shared origin of KBOs and Jupiter Trojans is the similar size distribution of the excited component of the Kuiper belt and the Trojans \citep{2009Icar..202..310M, 2014ApJ...782..100F}. A similar size distribution is, however, only a necessary but not sufficient condition to prove KBOs and Trojans have a common source. Complicating this prediction are the differences in the surface properties of KBOs and Trojans. Recent JWST spectroscopy has shown that KBOs have a rich surface chemistry, with water, carbon dioxide, methanol, and carbon monoxide ices frequently present \citep{2023PSJ.....4..130B, 2024NatAs.tmp..103D}. By contrast, Jupiter Trojans are spectroscopically neutral in the 1-4 micron range, with only weak organic and hydration features at around 3.0 microns \citep{2011AJ....141...25E,2016AJ....152..159B, 2019AJ....157..161W, 2024PSJ.....5...87W}. Resolving how two populations that ought to have the same origin have divergent spectral and photometric properties is necessary to prove the common origin of KBOs and Trojans -- an essential component of the Nice Model of the formation of the solar system. 

In this paper we find that the low-perihelion Kuiper belt objects (objects with $q < 30$ AU, $a > 30$ AU) have Trojan-like visible colors, and are a likely link between the Jupiter Trojans and Kuiper belt. In \autoref{sec:colors}, we discuss the colors of the various populations of the outer solar system, demonstrating how observations of the low-perihelion Kuiper belt place it apart from the rest of the Kuiper belt and Centaurs, and that the colors potentially bifurcate into those of the Jupiter Trojans. In \autoref{sec:dynamics} we investigate the two possible reasons for the unique colors of the low-$q$ Kuiper belt as opposed to the Centaurs -- different heating histories, or different source populations. Both of these possibilities are examined through the lens of dynamical simulations, finding that the difference in source populations between Centaurs and low-$q$ KBOs is the more likely reason for the more neutral colors of the low-$q$ KBOs.

\section{Colors of Outer Solar System Populations}
\label{sec:colors}

Objects in the outer solar system generally have featureless visible reflectance spectra ranging from neutral to red colors. To good approximation, these colors can be defined by a single parameter referred to as the spectral slope. The spectral slope is defined as the slope of the reflectivity at each observed wavelength, expressed in units of the percent change in the reflectivity over 100 nm, typically normalized at V band (550 nm). The slope can be calculated from photometric observations in different colors and filter systems, though it is primarily useful in the regime where the spectrum of objects tends to be linear, between 400-800 nm \citep{1986ApJ...310..937J, 2012A&A...546A.115H}. In order to obtain a full picture of the variety of color distributions in the many dynamical sub-populations that exist in the outer solar system, we combine four data sets. The first is from the Minor Bodies of the Outer Solar System (MBOSS) set of colors  \citep{2012A&A...546A.115H}, the second is the Colors of the Outer Solar System Origins Survey (Col-OSSOS) \citep{2019ApJS..243...12S}, the third is from the Dark Energy Survey (DES) \citep{2022ApJS..258...41B,2023ApJS..269...18B}, and the fourth is our own small survey of the colors of low-perihelion Kuiper belt objects, the technical details of which we describe in \autoref{sec:Observations}. We restrict ourselves to objects which have colors measured well enough that the uncertainty in the spectral slope is smaller than $\pm7\%$, equivalent to $g-i \pm 0.2$. We include only objects with absolute magnitude $H>7$ in order to remove larger objects whose colors are significantly different from smaller objects, hypothetically due to the geologic processing from accretional heating of larger objects \citep{2012AJ....143..146B}. Moreover, since we are comparing the colors of the Kuiper belt to the Trojans, we should be selecting only those KBOs which match the size range of the Jupiter Trojans, the largest of which are at $H \sim 7$ (see figure 6 of \citealt{2012A&A...546A.115H}). In total, we include 12 low-$q$ KBOs from OSSOS, 7 from MBOSS, 4 from DES, and 24 from our survey for a total of 47. For the Centaurs, 31 are taken from OSSOS, 38 from DES, none from DES, and one from our survey.
 \begin{figure}
    \centering
    \includegraphics[width = 8.5cm]{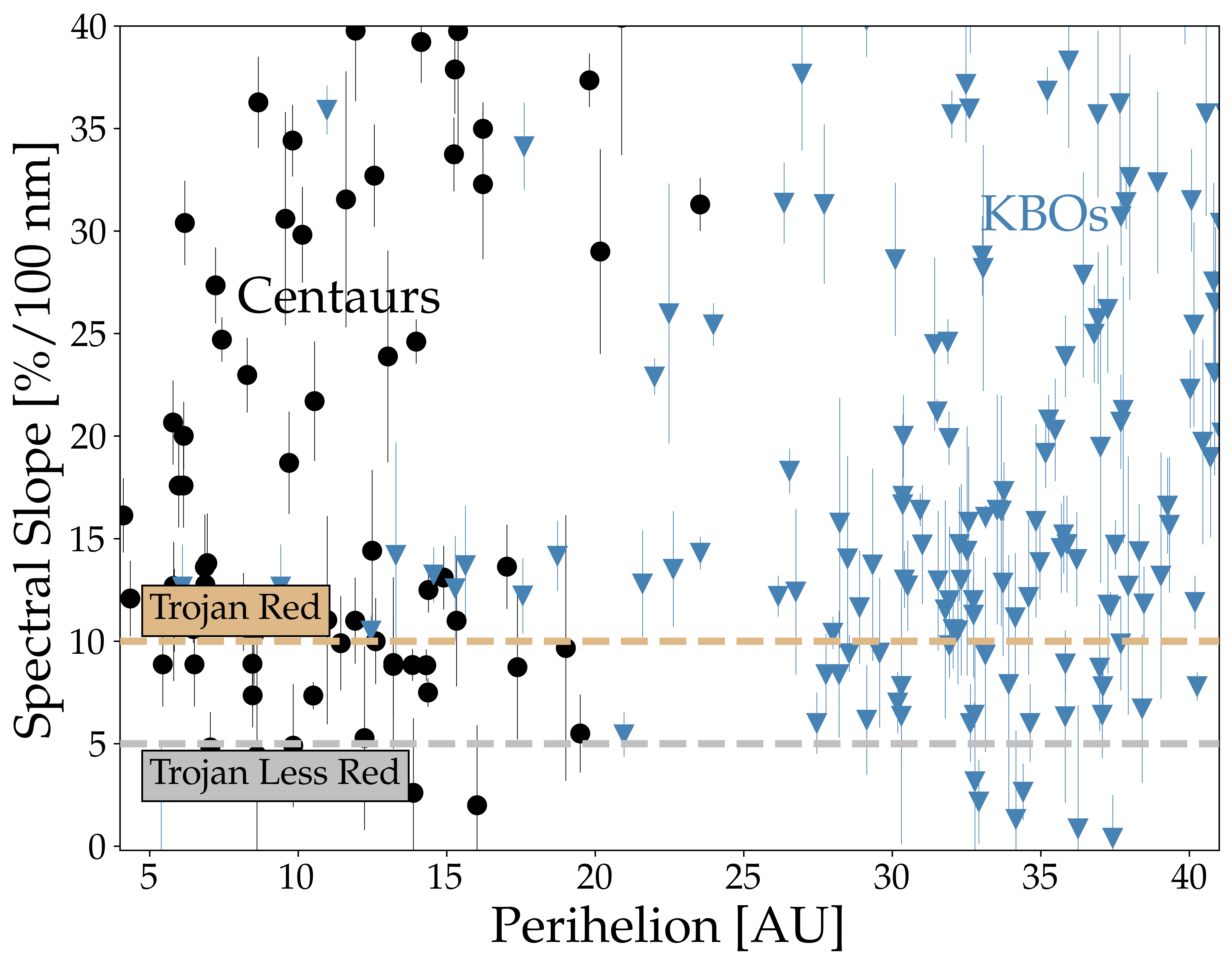}
    \caption{Colors and perihelia of the KBOs (blue; $a>30$ AU) and Centaurs (black; $a<30$ AU) taken from MBOSS and Col-OSSOS \citep{2012A&A...546A.115H,2019ApJS..243...12S}. The colors of the two Jupiter Trojan sub-populations are shown as dashed lines. While KBO and Centaur colors are statistically indistinguishable and significantly different from Jupiter Trojan colors, the colors of KBOs at low perihelia appear Trojan-like below $q<25$ au.}
    \label{fig:introplot}
\end{figure}
\begin{figure*}[t!]
    \centering
    \includegraphics[width = \textwidth]{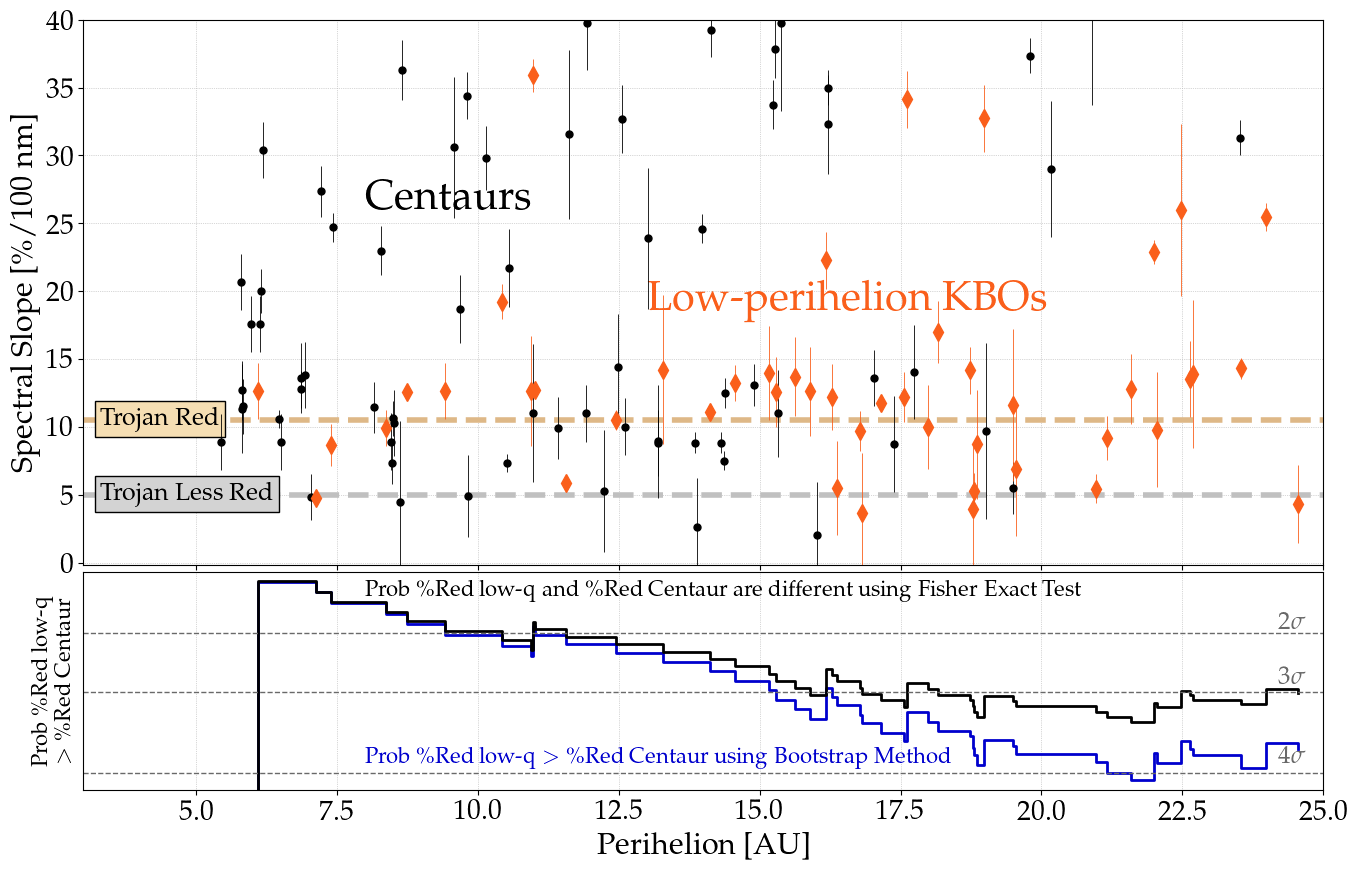}
    \caption{Statistical comparison of Centaur (black; $a<30$ AU, $q<30$ AU), and low-perihelion Kuiper belt (orange, $a>30$ AU, $q<30$ AU) colors, with Jupiter Trojan colors as dashed lines. The bottom panel shows the results of two tests of the difference in color of the Centaurs and low-perihelion KBOs: a one-sided Fisher exact test in black and a bootstrap method in blue. In both tests, objects are sequentially added to the analysis starting with the lowest perihelion objects. For any perihelion $15<q<25$ AU, the probability that the proportion of Red objects in the low-$q$ KBOs is greater from that of the Centaurs is more than 3$\sigma$ for the Fisher test, and nearly 4 $\sigma$ for the bootstrap method, indicating that the low-perihelion Kuiper belt is potentially a Trojan-like population in the outer solar system.}
    \label{fig:colors}
\end{figure*}

Kuiper belt objects with semi-major axis greater than Neptune's orbit have long been known to have bifurcated colors, split into a ``Very Red'' (KBO VR) population with slope greater than $20\%$ and a ``Red'' (KBO R) population with slope less than $20\%$, or $g-i = 1.15$ \citep{1998Natur.392...49T, 2002AJ....124.2279D, 2012A&A...546A.115H, 2017AJ....153..145W}. The Jupiter Trojan asteroids are also bifurcated in color, but into a ``Red'' (JT R) population, with a mean spectral slope of around 10\% or $g-i = 0.87$, and a ``Less Red'' (JT LR) population, with a mean spectral slope of around 5\% or $g-i = 0.74$ \citep{2011AJ....141...25E, 2016AJ....152...90W}. In this classification scheme, the Kuiper belt's ``Red'' colors fully encompass the range of Jupiter Trojan colors. 

Centaurs are former KBOs recently injected into the giant planet region and on unstable orbits with lifetimes typically between 100 kyr to 10 Myrs \citep{1997Icar..127...13L,1997Icar..127....1M,2003AJ....126.3122T,2010A&A...519A.112D}. For the purposes of this paper, we define Centaurs as having both perihelia and semi-major axes less than 30 AU. This population represents a transition from the cold temperatures of the KBOs to the warmer environments of the Jupiter Trojans. Though Centaurs have been substantially heated for many millions of years, no evidence exists that Centaurs are developing the surface colors of Jupiter Trojans -- their colors are statistically indistinguishable from those of the parent KBO population (see \autoref{fig:introplot}).

Previous publications have suggested color changes in Centaurs with heating, but, in each case, as more data have appeared, the statistical evidence has gone away (see \citealt{2012A&A...539A.144M, 2020tnss.book..307P}). As an aside, KBO surfaces clearly do evolve if they are heated enough. Jupiter family comets which originate from the Kuiper belt have darker surfaces than those typically seen on KBOs and have spectra generally unlike most KBOs, presumably as a result of significant ice sublimation over many perihelion passages \citep{2022PSJ.....3..251L, 2023PSJ.....4..208P}. Comet-like activity is seen in some of the closest Centaurs, though the correlation with surface colors remains unclear \citep{2009AJ....137.4296J, 2020ApJ...892L..38C}. JWST spectroscopy of the active comet 39P/Oterma, which is on a Centaur-like orbit, shows active CO$_2$ outgassing and a surface that has muted absorption features compared to KBOs observed with JWST \citep{2023PSJ.....4..208P}. Additionally, JWST spectroscopy by \cite{licandropreprint} has shown that the inactive Centaur Okyrhoe and Neptune co-orbital 2010 KR59 have spectra dissimilar to the larger KBO population, with broad 3.0 micron absorption features more similar to the Trojans shown in \cite{2024PSJ.....5...87W} rather than to the KBOs. The lack of color changes in the Centaurs further obfuscates the link between Trojans and the Kuiper belt. For example, hypotheses such as \cite{2016AJ....152...90W} suggested heating as the mechanism for the change from KBO R/VR colors to Jupiter Trojan LR/R colors, yet this transition is not observed in the colors of the Centaur population. 

A sub-population within the Kuiper belt that does, however, appear to have distinct colors is the low-perihelion Kuiper belt ($q<30$ AU, $a>30$AU). These objects are dynamically distinct from the Centaurs as they spend most of their orbits outside of Neptune, though both populations share similarly low perihelia. These low-$q$ KBOs appear to have a distinct set of colors from the Red/Very Red KBO and Centaur-like colors to Jupiter Trojan-like Red colors inside of $q<25$ AU, as shown in \autoref{fig:colors}. For perihelia between 25 and 30 au, there are both low-perihelion KBOs and resonant objects with large eccentricities that cross Neptune's orbit, so we exclude this region from our analysis.

We can statistically compare the difference in the two populations by computing the fraction of KBO ``Red'' versus ``Very Red'' objects in each population. Of the 70 Centaurs with $q<25$ AU in our sample, 28 are Very Red and 42 are Red, for a {Red fraction of $f_r=0.6 \pm 0.06$ (using the Wald interval for the binomial proportion confidence interval). Of the 47 low perihelion KBOs, 40 are Red, and only 7 are Very Red, for a Red fraction of $f_r=0.85 \pm 0.05$. Drawing 10,000 samples from the Centaur proportion of Red and Very Red objects and asking whether the low-perihelion KBOs has a larger proportion of Red objects than the Centaurs gives a 99.94\% confidence that the two populations have different proportions of each color for $q<25$ AU. Using even the very conservative one-sided Fisher exact test, the probability of the low perihelion KBOs having the fewer Red objects than the Centaurs is under $p=0.003$, or over a 99.7\% probability that their colors are different. In \autoref{fig:colors}, we iteratively repeat the Fisher test for each subset of the low-$q$ population by successively adding objects from lowest to highest perihelion. We find a more than three sigma probability that the low-$q$ KBOs have fewer Very Red objects than the Centaurs for any perihelion value above 17 AU. Moreover, our result for the Fisher exact test (or the more granular test in \autoref{fig:colors}) gives a confidence of over 99\% even when spectral slopes of 15-25\% are excluded, indicating that the difference in color between Centaurs and low-$q$ KBOs is robust and does not rely on a finely-tuned definition of the color bifurcation. 

\begin{figure*}
    \centering
    \includegraphics[width = 14cm]{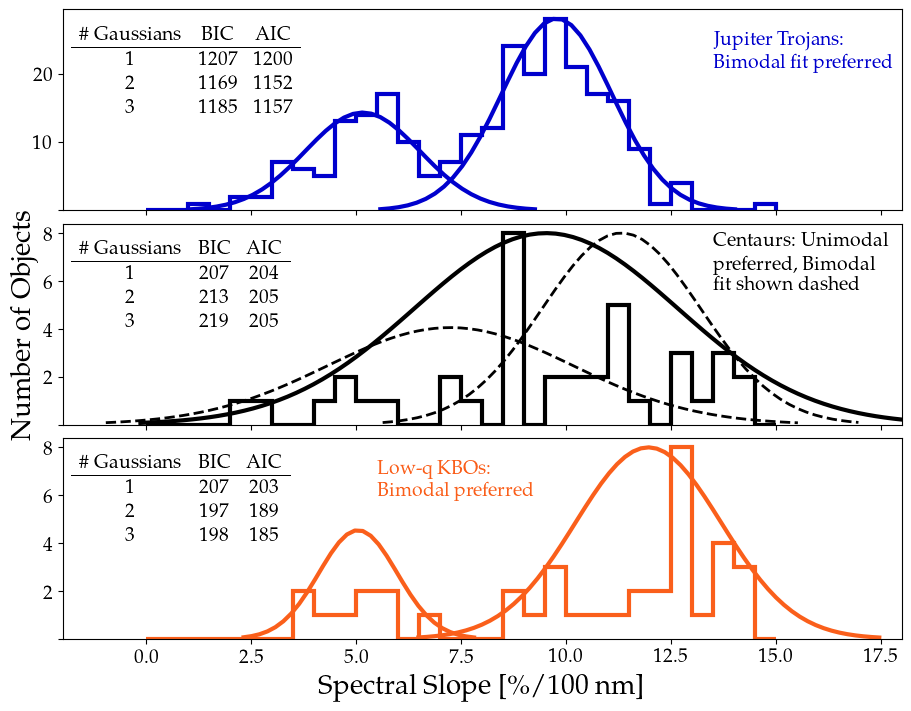}
    \caption{Gaussian mixture models for the spectral slopes of the Jupiter Trojans (top panel), Centaurs (middle panel), and low-perihelion Kuiper belt objects (bottom panel). Only objects with spectral slope $<15\%$, or those objects with Trojan-like slopes are included. For the Centaurs, a single Gaussian model is preferred to describe the data, with a double Gaussian fit shown in dotted lines for reference. By contrast, the spectral slopes of the low-$q$ KBOs are best described by a bimodal model, with a $\Delta$ AIC = 14 between the unimodal and bimodal model.}
    \label{fig:gmm}
\end{figure*}
\subsection{Testing the Trojan Bifurcation in Low-perihelion KBOs}
\label{sec:bifur}

We now test whether low-perihelion Kuiper belt objects have Trojan like colors by analyzing the distribution of their colors in the Trojan-like 0-15\% spectral slope range (KBO Red, or Jupiter Trojan LR/R colors).{The upper end of this range was set to be 15\% as that is the maximum spectral slope of Trojans reported in \cite{2016AJ....152...90W}. First, we check whether the color distribution of Red low-$q$ KBOs differs from that of similar spectral slope Centaurs. Selecting all Centaurs and low-$q$ KBOs with spectral slope $< 15\%$ and comparing the slope distributions with a Kolmogorov-Smirnov test, we find that the probability of them being drawn from the same distribution is at $p=0.058$. Although this result is just under a two-sigma significance, the 94\% probability of the two populations having a different distribution in the range of Jupiter Trojan colors invites us to test whether the low-$q$ KBOs or Centaurs are bifurcated similarly to the Jupiter Trojans. Using a gaussian mixture model, we test how many gaussians are preferred to describe the distribution of Centaur and low-$q$ KBO colors. For the low-$q$ KBOs, we find that a bifurcated model has a significantly lower Akaike Information criterion than a single gaussian model, with $\Delta$ AIC = 14 \citep{1974ITAC...19..716A}. For the Centaurs, the AIC is higher for any model with more than one Gaussian, indicating that a bifurcated model does not well-describe the slopes of the Centaurs. In \autoref{fig:gmm}, we show the result of mixture models for the spectral slopes of the Centaurs and low-$q$ KBOs along with the Jupiter Trojan colors from the Sloan Digital Sky Survey \citep{2014AJ....148..112W}. We report both the AIC and BIC for the one, two, and three gaussian mixture models for all three populations. The two means of the mixture model for the slopes of the slope $< 15\%$ low-$q$ KBOs are close to the two colors of the Jupiter Trojan population, implying that the ``Red'' low-$q$ KBOs have a similar color distribution to the Jupiter Trojans. A factor that also suggests the similarity between low-$q$ KBOs and Trojans is that the ratio of objects with Trojan Red colors to Trojan Less Red colors seems to be replicated in the low-$q$ KBO population. However, a notable inconsistency between the Trojan and low-$q$ KBO colors is that redder component of the two sub-populations in the low-$q$ KBOs is slightly redder than the mean colors of the ``Trojan Red'' subpopulation, as seen in the right panel of \autoref{fig:gmm}. A similar effect was found in \cite{2017AJ....153...69W}, who examine the Hilda asteroid population and find that their colors are bifurcated similarly to the Trojans, but with the mean colors shifted towards slightly more neutral values. The Hildas lie between the asteroid belt but interior to Jupiter in the 3:2 mean motion resonance with Jupiter at 4 AU, and are thought to have a common origin story with the Trojans \citep{Marsset_2014, 2017AJ....153...69W}. The successive shifts in the color bifurcation from the Trojan-like component of the low-$q$ KBOs to the Jupiter Trojans to the Hildas indicates that spectral slopes could become slightly more neutral with decreasing heliocentric distances. Our evidence for the Trojan-like bifurcation among more neutrally-colored (Red, slope $<15\%$) low-$q$ KBOs suggests there exists a subset of objects in the Kuiper belt with Jupiter Trojan-like colors, giving an observational link between the two populations.

\subsection{Highly-inclined Objects}
An interesting subset of objects in the Centaurs and the low-perihelion KBOs are those that have high or retrograde inclinations. The dynamical origin of $i>50^\circ$ objects remains an open question \citep{2019PhR...805....1B}. Simulations of the Nice model fail to produce the high-inclination ($i>50^\circ$) objects we observe in the Kuiper belt \citep{2015AJ....150...73N}, such as 2015 BP$_{159}$ or any of the numerous $i>90^\circ$ Centaurs and low-perihelion KBOs \citep{2009ApJ...697L..91G, 2018AJ....156...81B}; simply put, the disk-like nature of a protoplanetary disk likely precludes the formation of any sizeable reservoir of highly-inclined objects. Moreover, starting from the currently observed Kuiper belt, there is no way to produce retrograde objects with $q>5.2$ AU with the four giant planets \citep{2015AJ....150...73N}. All of the 14 low-$q$ KBOs and 3 Centaurs with $i>50^\circ$ have remarkably similar colors, with spectral slopes between 5\% and 20\%. If we repeat our statistical comparison of the Centaurs and low-$q$ KBOs without these inclined objects, the significance of the difference between Centaurs and low-$q$ KBOs lessens from more than 3$\sigma$ to over 2$\sigma$, largely due to the decreased sample size. Similarly, evidence for the bifurcation in \autoref{fig:gmm} weakens, with a less significant $\Delta$AIC $\sim 7$. However, these retrograde objects are a key feature of the low-$q$ KBOs and the Centaurs -- our current inability to assess the dynamical pathways of these $i>60^\circ$ objects does not reduce the significance of the difference in colors of the Centaurs and low-$q$ KBOs. Therefore, our dynamical analysis in the remainder of this paper excludes the high inclination ($i>50^\circ$) objects from our sample, as these objects cannot be replicated in dynamical simulations of the Kuiper belt and the four gas giants \citep{2017AJ....154..229B,2019PhR...805....1B}.

\section{Origin of Trojan-like colors}
\label{sec:dynamics}

We present two possible testable hypotheses for why the colors of the low-perihelion Kuiper belt are distinct from the Centaurs. The first hypothesis is that the average low-$q$ KBO is heated longer than the average Centaur, causing their surfaces to undergo the chemical changes that produce Jupiter Trojan colors, resulting in the population-level color difference we observe. The second hypothesis is a purely dynamical one -- the low-$q$ KBOs and Centaurs have distinct source populations with distinct colors, which are then carried over to the two excited populations. To test these hypotheses, we need to obtain the distribution of possible Kuiper belt origins and dynamical lifetimes of our sample of Centaurs and low-$q$ KBOs, and determine whether the populations have significantly different heating histories or source populations. 

\begin{figure*}
    \centering
    \includegraphics[width = 18cm]{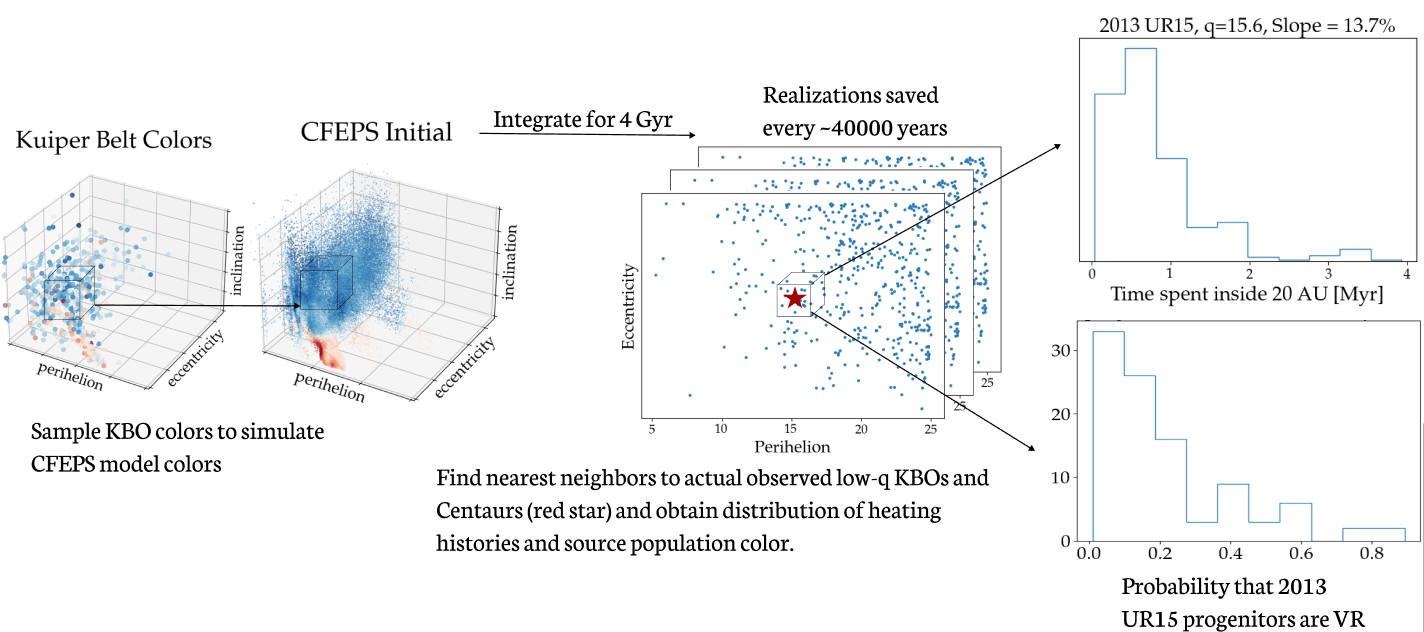}
    \caption{Schematic of the method used in \autoref{sec:dynamics} for tracking the source and dynamical evolution of low-$q$ KBOs and Centaurs. We begin with the Canada–France Ecliptic Plane Survey (CFEPS) L7 simulated Kuiper belt reference population \citep{2011AJ....142..131P}, integrating it forward for 4 Gyr with the four giant planets. These integrations produce around 150,000 realizations of the $q<25$ AU component of the Kuiper belt. Simultaneously, we find the color distribution for each CFEPS object based on the known colors of nearby Kuiper belt objects, using a K-nearest neighbor algorithm. During the integration, we also track the heating history (time spent within 20 AU) of all of the synthetic objects. Using all of these snapshots of synthetically generated low-$q$ KBO and Centaur populations, we then compute statistics on real objects that we have colors for by drawing the closest objects from our simulations. As a result, we obtain distributions on both the possible heating histories and color as a function of original Kuiper belt source for the Centaurs and low-$q$ KBOs from \autoref{sec:colors}.}
\label{fig:diagram}
\end{figure*}

Dynamical histories of unstable objects are impossible to directly ascertain as backwards integration gives radically different results from infinitesimally small changes to initial conditions. Even starting with thousands of clones for each object and integrating backwards does not correctly link observed Centaurs and low-$q$ KBOs with the Kuiper belt. This method starts with a problematic assumption that these clones are all equally likely to appear from the scattering of Kuiper belt objects. Instead, we create a forward model, integrating a sample Kuiper belt for four billion years to see what differences exist in the dynamical behavior of the bodies that produce Centaurs and low-$q$ KBOs. 

Our forward model begins with the Canada–France Ecliptic Plane Survey (CFEPS) L7 simulated Kuiper belt reference population \citep{2011AJ....142..131P}. We first integrate the reference population with the four giant planets using the \texttt{MERCURY} integrator \citep{1999MNRAS.304..793C}. We exclude objects that are initially low-$q$ KBOs (and thus have an uncertain source) to avoid contaminating our sample of the excited population. We use a hybrid changeover radius of 6 Hill radii, a timestep of 150 days, and an ejection distance of 10000 AU. Every 10 million days (around 25,000 years), we record instantaneously whether each object fits our definition of a Centaur or low-$q$ KBO and its initial orbital elements. Over four billion years, we obtain 150,000 snapshots of our populations of interest that contain 25.1 million synthetic low-$q$ KBOs and 3.1 million Centaurs. With millions of simulated Centaurs and low-$q$ KBOs each with a full dynamical history, we can begin to test whether a difference in heating or difference in source population is responsible for the distinct color distribution of the Centaurs and low-$q$ KBOs. 

We cannot, however, compare all simulated low-$q$ KBOs to all simulated Centaurs to answer our hypotheses. To explain the colors we observe in the specific sample from \autoref{sec:colors}, we must select simulated low-$q$ KBOs and Centaurs that are similarly distributed to the real low-$q$ KBOs and Centaurs. For each low-$q$ KBO and Centaur in the photometric data from \autoref{sec:colors}, we compute statistics on its 1,000 nearest synthetic neighbors. Nearest neighbors are determined using the Euclidean distance in eccentricity/inclination/perihelion space, where we have normalized each parameter to be between zero and one. A schematic of this method is shown in \autoref{fig:diagram}.

\subsection{Heating Hypothesis}
To test the heating hypothesis, we track how much insolation each synthetic object has experienced using a simple metric: time spent within 20 AU. Specifically, we track the time a given object is inside of 20 AU within its orbit, not the time an object has a perihelion of less than 20 AU. This specific threshold is an arbitrary one, and is not due to any sublimation line -- repeating our analysis for other thresholds such as 10 or 15 AU produces similar results. By selecting the 1,000 synthetic objects closest to each real Centaur and low-$q$ KBO in our sample, we can obtain a statistical distribution of the time spent inside of 20AU for each object we have colors for. The left panel of \autoref{fig:heating} shows the median amount (with one sigma error bars) of time spent inside of $r<20$ AU as a function of current perihelion distance for the Centaurs or low-$q$ KBOs used in our photometric analysis in \autoref{sec:colors}. For Centaurs and low-$q$ KBOs, the typical time spent within 20 AU is 1 to 10 million years, with the low-$q$ KBOs past 15 AU spending more time within 20 AU than the average Centaur.

\begin{figure*}
    \centering
    \includegraphics[width = 8.5cm]{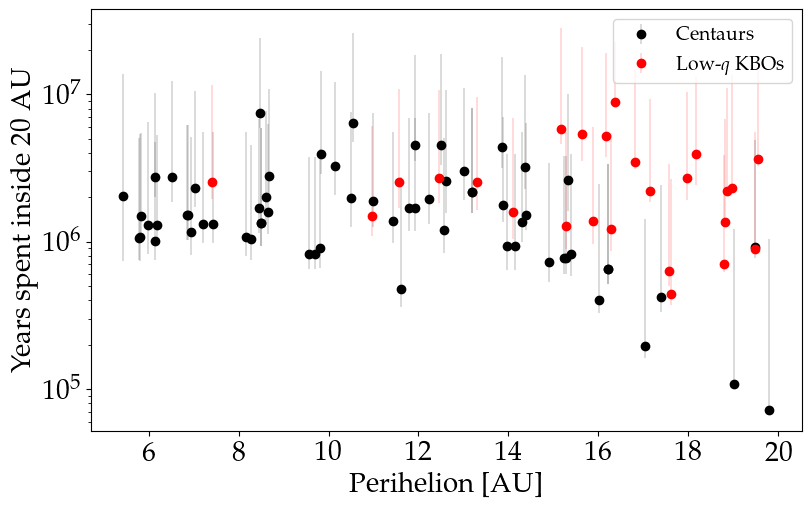}
    \includegraphics[width = 8.5cm]{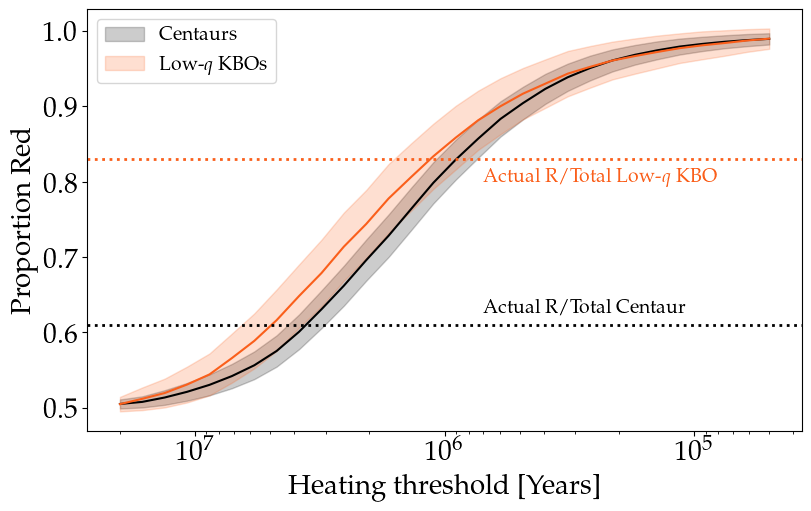}
    \caption{\textit{Left panel}: The median total years spent within 20 AU of realizations of simulated objects with orbital elements similar to the observed low-$q$ KBOs (in orange dots) and Centaurs (in black dots). The Centaurs spend a similar amount of time within 20 AU as the low-$q$ KBOs. \\ \textit{Right panel:} Simulation of the proportion of Red objects in the two populations (starting from around the KBO Red proportion), assuming heating is responsible for changing Very Red objects into Red ones. The difference between Centaurs and low-$q$ objects in their Red/total ratio never approaches that of the photometric data. Heating is therefore not sufficient to create the observed difference between the two populations.}
    \label{fig:heating}
\end{figure*}

Having found the distribution of time spent within 20 au for each real Centaur and low-$q$ KBO using the synthetic population generated from the CFEPS simulations, we can evaluate whether the heating histories of the two populations produce the observed distinct colors. In order to determine whether distinct heating histories can reproduce the colors of the two populations, we test whether there exists a critical threshold for time spent within 20 au for the change from Very Red to Red colors that would create a significantly different proportion of Red to Very Red objects in the low-$q$ KBOs as opposed to Centaurs. For each real low-$q$ KBO and Centaur in the left panel of \autoref{fig:heating}, we assign a base 50\% chance of the object being Red. Starting from a threshold of 20 million years of heating to transition from Very Red to Red, we determine the probability of each object having transitioned to Red colors, given its distribution of time spent within 20 AU. In other words, for each real object, we determine its probability of being Red based on how many of its synthetic look-alikes have spent more time within 20 au within each given time threshold. We then simulate 10,000 draws for the colors of all the objects, thus producing 10,000 simulations of the colors of Centaur and low-$q$ KBOs populations. We then get the median proportion of Red objects in each population as a function of heating threshold for transitioning between the two colors, which is shown in the right panel of \autoref{fig:heating}. Regardless of the choice of the minimum amount of time it takes for an object to change color, heating of low-$q$ KBOs and Centaurs results in the low-$q$ KBOs having only slightly fewer Very Red objects than the Centaurs, not distinct enough to match the real proportions. In the 10,000 simulations performed at each heating threshold, at most 4\% of the simulations yield a color difference between the two populations that is more than the observed one. All heating thresholds outside the 1-3 Myr range result in all of the simulations yielding a difference in the proportion of Red objects between the two populations that is less than the one observed. It is therefore unlikely that irradiation of surfaces results in the difference between the colors of low-$q$ KBOs and Centaurs.

\begin{figure*}
\centering
    \includegraphics[width = 16cm]{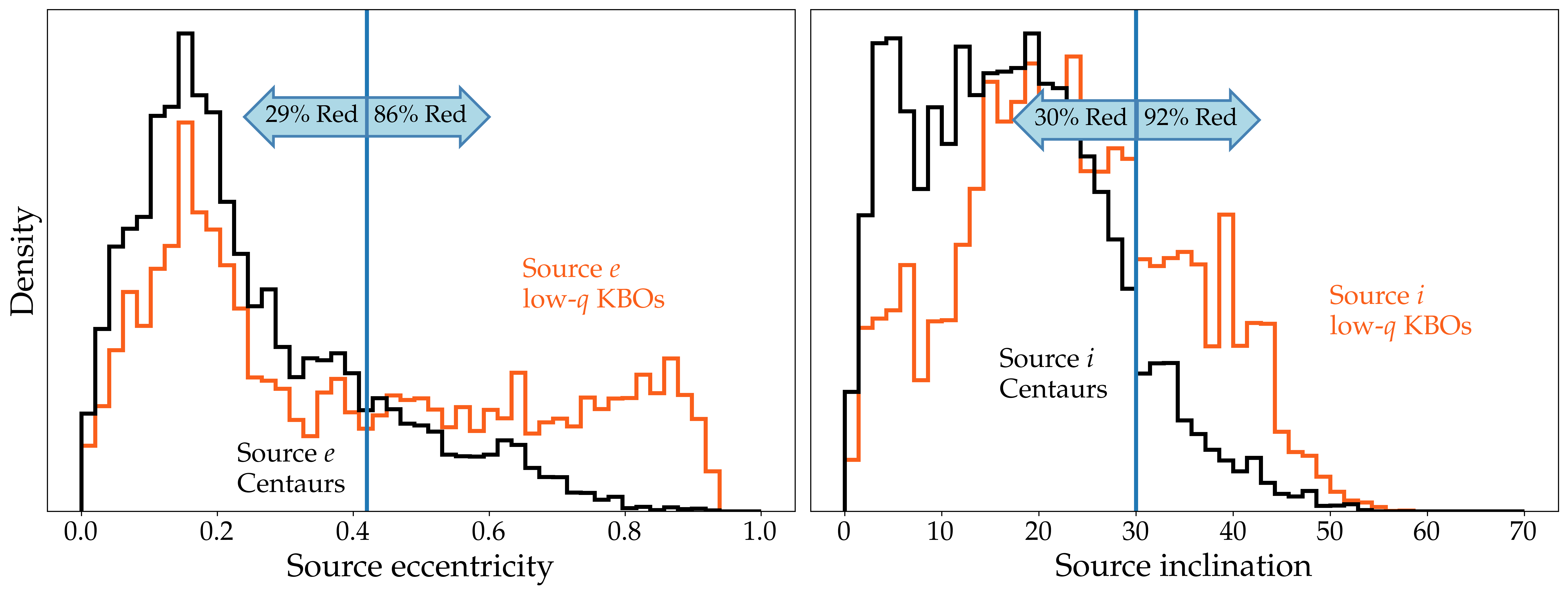}
    \caption{Distribution of the eccentricity and inclination of source objects that become dynamically similar to known low-q KBOs or Centaurs at any point in out 4 Gyr simulations. The low-q Kuiper belt has a unique source from objects with $e > 0.6$ or $i > 40^\circ$. Objects with those dynamical characteristics in the Kuiper belt tend to not have very red colors, thus the origin of low-q KBOs in these dynamically hot populations causes the population to have significantly more Red (fewer Very Red) objects. A similar analysis of the rarity of VR objects in the scattered disk was presented in \cite{2021AJ....162...19A}.}
    \label{fig:dynamics}
\end{figure*}

Heating and stability are a non-trivial result of dynamical interactions, but our modelling suggests that heating is insufficient as an explanation for the more neutral colors of low-$q$ KBOs as compared to the Centaurs. Other, more physical, means of modelling heating produce the same result. Computing the solar flux accumulated over the lifetimes of these objects after they cross inside of 30 AU produces a nearly identical result to time spent within 20 AU. We use time spent within 20 AU as it is the most conservative metric for heating given the question we are testing -- Centaurs are on average at lower perihelia than low-$q$ KBOs, thus, if they spend a similar amount of time within 20 AU, low-$q$ KBOs are heated less. We note that there are indeed spectral changes in the infrared observed on comets that are a result of surface processing, such as those that occur to the 1.5, 2.0, 3.0 micron water features, or the CO$_2$ absorption feature \citep{2023PSJ.....4..208P, licandropreprint}. Rather, we find evidence that surface processing on timescales equal to or less than the dynamical lifetimes of these excited objects cannot reproduce the color difference in the visible between Centaurs and the low-$q$ KBOs. Thus, the origin of Jupiter Trojan-like spectral slopes in the visible in the low-$q$ KBO population has to be addressed by comparing the dynamical sources of the two populations.

\subsection{Source Population Hypothesis}

Since heating is unlikely to be responsible for the difference in color between Centaurs and low-$q$ KBOs, we test whether the color distribution in the $q>30$ AU Kuiper Belt, which is the source population for the Centaurs and low-$q$ KBOs, can account for the distinct colors of the two populations. We test whether there are differences in source location of the low-$q$ KBOs and Centaurs from which they inherit their distinct colors. We show the distribution of source eccentricities and inclinations for Centaurs and low-$q$ KBOs in \autoref{fig:dynamics}. The source eccentricities and inclinations are determined by selecting synthetic low-$q$ KBOs and Centaurs near the observed ones and tracing them back to their origin in CFEPS, recording the starting eccentricity and inclination. Despite objects commonly transitioning between these two dynamical classes, the two populations are unevenly sampling the same Kuiper belt source. Centaurs preferentially come from the low-$e$, low-$i$ part of the Kuiper belt, while low-$q$ KBOs are more likely than Centaurs to come from the high-eccentricity scattered disk or high-inclination populations such as the most excited of the hot classicals and Plutinos. The more excited populations in the Kuiper belt are known to have fewer VR objects, as shown in \autoref{fig:dynamics}. For objects with eccentricity above 0.4, around 14\% are VR, whereas 71\% are VR for lower eccentricities. Similarly, among the objects with inclinations above 30$^\circ$, only 8\% are VR. Given that the Centaurs draw more frequently from the less excited population that has significantly more Very Red objects than the excited population, we can test whether these differences are enough to account for their different observed source colors.

Having found that low-$q$ KBOs and Centaurs have differences in their source populations, we can test whether this difference is enough to reproduce the color difference of the two populations. First, we need to determine what is the probability of each synthetic object being Red or Very Red as a consequence of its Kuiper belt source in the CFEPS population. Since the CFEPS population is a model of the observed Kuiper belt, we can use observed colors from MBOSS, Col-OSSOS, and DES \citep{2012A&A...546A.115H,2019ApJS..243...12S, 2022ApJS..258...41B,2023ApJS..269...18B} to get a likelihood that a given CFEPS reference population object is R or VR. We exclude Neptune Trojans, $H<7$, and Slope error $<7\%$ objects from our Kuiper belt color sample. For each object in CFEPS, we determine its color using a K-nearest neighbors regression on the observed colors from the aforementioned sources. In other words, the color of each object in CFEPS is given by the distance-weighted average of nearby observed Kuiper belt colors. This distance is the Euclidean distance between the CFEPS object and any given object in the MBOSS/Col-OSSOS/DES sample, where inclination, eccentricity, and perihelion are normalized to be between 0 and 1, such that each dynamical parameter is equally weighted. For example, a synthetic low-$q$ KBO that is sourced from the classical Kuiper belt is more likely to be Very Red than one coming from the Scattered disk.

Now that we have the likelihood of an object being R/VR for each synthetic object, we can determine the probability that our real objects are R/VR as described in the beginning of this section. We take the nearest 1000 simulated neighbors to each real low-$q$ KBO and Centaur, and calculate a distance-weighted average of the R/VR probabilities of the synthetic objects. As a result, for each low-$q$ KBO and Centaur in our photometric sample, we have the corresponding probability of any given object being R/VR based on its Kuiper belt source. 

\begin{figure}
    \centering
    \includegraphics[width=8.5cm]{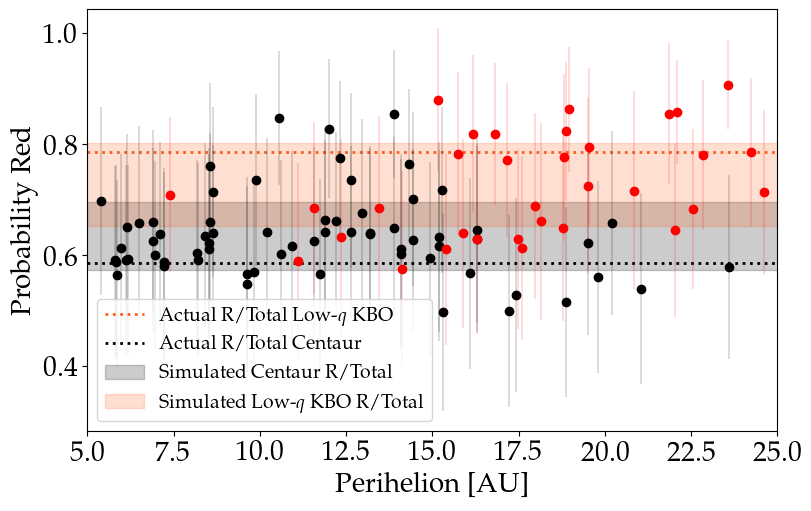}
    \caption{Scatterplot of the median probability that a low-$q$ KBO (orange) or Centaur (black) from our photometric sample is Red given its dynamical origin in the Kuiper belt. This probability is determined using the method in \autoref{fig:diagram}, wherein each real Centaur and low-$q$ KBO is linked to possible Kuiper belt progenitors, whose colors are simulated using a combination of available photometric data. Dashed lines show the true proportions in observed Centaurs and low-$q$ KBOs with $q<25$ AU, while the shaded areas show the simulated proportion of Red objects with one sigma error. The simulated proportions are consistent with the data proportions for the number of Red objects in the Centaurs and low-$q$ KBOs.}
    \label{fig:dynorigin}
\end{figure}

Finally, to test our hypothesis, we compute the proportion of the low-$q$ KBOs and Centaurs that we expect to be Red, similarly to the statistic obtained in \autoref{sec:colors}, where $58\pm 6\%$ of Centaurs and $78\pm 7\%$ of low-$q$ KBOs (now excluding the retrograde objects from the color sample) were found to be Red. Given that we have the probability of each real Centaur and low-$q$ KBO being red, we can simulate 10,000 draws for the color of each object, thus producing 10,000 simulations of the colors of Centaur and low-$q$ KBOs populations. In \autoref{fig:dynorigin}, we show the probability that each Centaur and low-$q$ KBO used in our photometric analysis has Red/Very Red colors based on its Kuiper belt source. The shaded regions show the one sigma range for the population proportion of Red objects for the two populations, while the dashed lines show the true proportions from \autoref{sec:colors}. We find that the proportion of Red low-$q$ KBOs derived from propagating the colors of their source population is 0.73 $\pm$ 0.07 (where the observed value for the proportion of Red $q<25$ au low-$q$ KBOs is 0.78 $\pm$ 0.07), while for Centaurs the Red proportion is 0.63 $\pm$ 0.06 (where the observed proportion is 0.58 $\pm$ 0.06). 

While the data proportions of Red objects for both Centaurs and low-$q$ KBOs are within one sigma error bars of the simulated proportions, there is overlap between the simulated proportions for both sets of objects. This indicates that while we can match the Centaur proportion of Red objects or the low-$q$ KBO proportion of Red objects, we do not necessarily match the difference between the two populations of 20\%. Of our 10,000 simulations of the Centaur and low-$q$ KBO populations, just over 17\% have a difference in Centaur and low-$q$ KBO Red proportion of 20\% or more.  While this result indicates it is possible that source population differences explain color differences and that this hypothesis is a more likely one than the heating hypothesis, we cannot assert source population differences as the complete explanation without more evidence. One likely possibility for our model not explaining the full difference in colors of the Centaurs and low-$q$ KBOs is that CFEPS does not include enough high eccentricity and high inclination objects. The scattered disk in the CFEPS model is known to be too small \citep{2022ApJS..258...41B}, and more current models, such as those being developed from the OSSOS survey include notably more SDOs and higher-order resonant objects (Kavelaars, Personal communication, 2023). If we increase the number of both $e>0.8$ and $i>40^\circ$ objects at the start of the CFEPS simulation by an order of magnitude (simply by duplicating them), we find that in over 35\% of our simulations we match the observed difference between the two populations.

Stepping back to the bigger question, the link between colors in the scattered disk, the low-perihelion component of the Kuiper belt, and the Jupiter Trojans provides observational support for the Nice model. According to the Nice model, the scattered disk is sourced from the inner part of the outer planetesimal disk (24-30AU) \citep{2020AJ....160...46N}, similarly to the Jupiter Trojans, which are the inner disk planetesimals that are scattered and then captured by Jupiter during its final instability \citep{2013ApJ...768...45N}. The scattered disk indeed has more neutral colors relative to the rest of the Kuiper belt, similar to the Jupiter Trojans. Our observations of the low-perihelion component of the Kuiper belt find it to have the same preferentially ``Red'' colors. Then from dynamical simulations, we find that the low-$q$ KBOs preferentially originate from the scattered disk, selecting mostly those objects which were part of the inner part of the planetessimal disk, from which Jupiter Trojans originate. Thus, the Trojan-like colors of the low-$q$ KBOs suggest that the scattered disk and Jupiter Trojans have a common initial Nice model origin in the component of the planetesimal disk closest to the sun.

\section{Discussion}
\label{sec:discussion}

Combining our photometric observations on Keck and Palomar with all available data from MBOSS \citep{2012A&A...546A.115H}, Col-OSSOS \citep{2019ApJS..243...12S}, and DES \citep{2022ApJS..258...41B,2023ApJS..269...18B}, we have isolated the low-perihelion Kuiper belt as a population with Jupiter Trojan-like colors in the outer solar system. The measured colors show that these KBOs with $q<30$ AU and $a>30$ AU have a proportion of Red to Very Red objects more than 3$\sigma$ distinct from Centaurs, which have similar perihelion distances. Critically, we note that the low-perihelion KBOs display a similar bifurcation in colors to the Jupiter Trojans, with a slightly redder color, mimicking how the Hildas, interior to the Trojans, are also bifurcated, but with a slightly bluer set of colors. 

More importantly, we have found observational evidence for the Nice model, by linking the Jupiter Trojans and Scattered Disk Objects, which according to the Nice model, have a common source in the inner disk. The low-$q$ KBOs share the more neutral colors of the Scattered Disk and Jupiter Trojans, while having a distinct source in the scattered disk, unlike the Centaurs, which are sourced from the more stable parts of the Kuiper belt. We also demonstrate that these colors cannot have arisen from heating of the low-perihelion population, as similar-perihelion KBO-colored Centaurs and Trojan-colored low-$q$ KBOs have heating histories which do not result in the drastic difference in color shown in \autoref{fig:colors}. Instead, we show that the difference in colors of Centaurs and low-$q$ KBOs is plausibly explained by the two populations having different Kuiper belt progenitors. The low-$q$ KBOs are more frequently sourced from the scattered disk and high-inclination population than the Centaurs. Our results indicate that the difference in source populations is preferable to heating as an explanation for the distinct proportions of Red objects in the low-$q$ KBOs and Centaurs. Further improvements to both the simulated models of the Kuiper belt as well as our knowledge of its colors should help fully resolve the question of the origin of the difference in Centaur and low-$q$ KBO colors.


\appendix
\section{Observational Details}
\label{sec:Observations}

\begin{table*}
    \caption{Palomar Observations.}
    \begin{tabular}{lllll}
    \hline
    Name & Mean $\mathrm{m_g^1}$ & Mean $\mathrm{m_i^2}$ & Spec. Slope$^3$ & Obs. Date and UT$^4$\\ 
     & & & (\%/100 nm) & \\ \hline
    2004 TE282 & 24.11 $\pm$ 0.11 & 22.98 $\pm$ 0.07 & (18.4 $\pm$ 5.1)\% & 2018 Sep. 23 10:48\\ 
    2009 MS9 & 22.19 $\pm$ 0.04 & 21.23 $\pm$ 0.04 & (12.7 $\pm$ 0.3)\% & 2018 Sep. 23 09:14\\
    2010 JG210 & 22.56 $\pm$ 0.01 & 21.62 $\pm$ 0.01 & (11.7 $\pm$ 0.3)\% & 2019 July 27 05:37\\
    2011 LZ28 & 22.17 $\pm$ 0.17 & 21.38 $\pm$ 0.14 & (6.9 $\pm$ 4.4)\% & 2020 Sep. 18 07:10\\
    2013 KZ18 & 21.98 $\pm$ 0.06 & 21.13 $\pm$ 0.15 & (8.8 $\pm$ 4.1)\% & 2020 Sep. 18 07:32\\
    2014 JF80 & 21.71 $\pm$ 0.09 & 21.01 $\pm$ 0.08 & (4.0 $\pm$ 4.9)\% & 2020 Sep. 18 05:25\\
    2014 LN28 & 21.04 $\pm$ 0.08 & 20.19 $\pm$ 0.04 & (8.7 $\pm$ 3.3)\% & 2020 Sep. 18 04:42\\
    2014 QV441 & 22.06 $\pm$ 0.06 & 21.31 $\pm$ 0.03 & (5.3 $\pm$ 1.3)\% & 2020 Sep. 18 09:45\\
    2014 SK349 & 22.92 $\pm$ 0.14 & 22.01 $\pm$ 0.12 & (11.3 $\pm$ 0.5)\% & 2020 Sep. 18 10:31\\
    2014 UG241 & 23.08 $\pm$ 0.11 & 22.12 $\pm$ 0.11 & (12.5 $\pm$ 3.0)\% & 2020 Sep. 18 10:12\\
    2014 UT114 & 21.96 $\pm$ 0.07 & 20.99 $\pm$ 0.08 & (12.6 $\pm$ 3.3)\% & 2018 Sep. 23 10:31\\
    2014 WD536 & 23.10 $\pm$ 0.08 & 22.17 $\pm$ 0.10 & (11.7 $\pm$ 5.5)\% & 2020 Sep. 18 09:55\\
    2014 WR509 & 22.66 $\pm$ 0.06 & 21.75 $\pm$ 0.13 & (10.6 $\pm$ 3.4)\% & 2020 Sep. 18 10:50\\
    2015 KZ120 & 20.52 $\pm$ 0.06 & 19.64 $\pm$ 0.02 & (9.9 $\pm$ 1.3)\% & 2019 July 27 05:53\\
    2017 RG16  & 21.61 $\pm$ 0.02 & 20.88 $\pm$ 0.01 & (4.8 $\pm$ 0.2)\% & 2020 Sep. 18 07:22\\
    \label{tab:obspal}
    \end{tabular}
    \begin{tablenotes}
    \item 1) Average apparent $g$-band magnitude. 2) Average apparent $i$-band magnitude. 3) Spectral slope is calculated from successive $g-i$ color measurements (where the slope is normalized to V-band following \cite{2012A&A...546A.115H}), and will not precisely match the slope calculated from the average magnitudes in each band. 4) Time is of the first observation in the set.
    \end{tablenotes}
\end{table*}

\begin{table*}
    \caption{Keck Observations}
    \begin{tabular}{lllll}
    \hline
    Name & $\mathrm{m_g^1}$ & $\mathrm{m_i^2}$ & Spec. Slope$^3$ & Obs. Date\\ 
     & & & (\%/100 nm) & \\ \hline
    2005 ER318 & 23.44 $\pm$ 0.01 & 22.62 $\pm$ 0.03 & (7.7 $\pm$ 0.8)\% & 2022 April 28 08:53\\
    2007 BP102 & 23.97 $\pm$ 0.07 & 22.97 $\pm$ 0.09 & (14.0 $\pm$ 3.5)\% & 2021 May 15 08:07\\
    2008 KV42 & 23.47 $\pm$ 0.03 & 22.61 $\pm$ 0.03 & (9.2 $\pm$ 1.6)\% & 2021 May 15 12:58\\
    2011 FY9 & 23.16 $\pm$ 0.04 & 22.16 $\pm$ 0.10 & (13.9 $\pm$ 3.5)\% & 2021 May 15 11:21\\
    2012 GU11 & 23.80 $\pm$ 0.03 & 22.71 $\pm$ 0.05 & (17.0 $\pm$ 2.3)\% & 2021 May 15 11:06\\
    2012 GX17 & 23.57 $\pm$ 0.05 & 22.87 $\pm$ 0.14 & (3.7 $\pm$ 4.4)\% & 2021 May 15 07:54\\
    2013 FJ28 & 22.77 $\pm$ 0.16 & 21.84 $\pm$ 0.06 & (11.6 $\pm$ 5.6)\% & 2021 May 15 07:37\\
    2013 LU28 & 19.42 $\pm$ 0.01 & 18.46 $\pm$ 0.01 & (12.6 $\pm$ 0.5)\% & 2022 April 28 07:55\\
    2013 ME14 & 23.84 $\pm$ 0.05 & 22.34 $\pm$ 0.07 & (32.7 $\pm$ 2.5)\% & 2021 May 15 13:38\\
    2014 BF70 & 21.26 $\pm$ 0.03 & 20.49 $\pm$ 0.01 & (5.8 $\pm$ 0.4)\% & 2022 April 28 08:51\\
    2014 LM28 & 23.18 $\pm$ 0.07 & 22.30 $\pm$ 0.09 & (9.7 $\pm$ 1.5)\% & 2022 April 28 12:56\\
    2014 LN28 & 21.47 $\pm$ 0.06 & 20.52 $\pm$ 0.02 & (12.2 $\pm$ 2.4)\% & 2021 May 15 14:18\\
    2014 XO40 & 21.11 $\pm$ 0.04 & 19.88 $\pm$ 0.04 & (22.3 $\pm$ 2.1)\% & 2022 April 28 09:59\\
    2014 XQ40 & 22.31 $\pm$ 0.09 & 21.42 $\pm$ 0.04 & (10.0 $\pm$ 3.1)\% & 2022 April 28 10:12\\
    2015 BE568 & 22.91 $\pm$ 0.17 & 22.11 $\pm$ 0.07 & (6.9 $\pm$ 4.9)\% & 2021 May 15 07:43\\
    2017 CX33 & 23.23 $\pm$ 0.06 & 22.08 $\pm$ 0.02 & (19.2 $\pm$ 1.3)\% & 2021 May 15 05:37\\
    2017 FO161 & 23.50 $\pm$ 0.09 & 22.15 $\pm$ 0.10 & (27.0 $\pm$ 2.3)\% & 2021 May 15 06:02\\
    2017 KZ31 & 23.04 $\pm$ 0.12 & 22.08 $\pm$ 0.06 & (12.6 $\pm$ 4.1)\% & 2022 April 28 05:27\\
    2017 YG5 & 22.45 $\pm$ 0.06 & 21.61 $\pm$ 0.02 & (8.6 $\pm$ 1.6)\% & 2022 April 28  08:48\\
    \label{tab:obskeck}
    \end{tabular}
    \begin{tablenotes}
    \item 1) Apparent $g$-band magnitude. 2) Apparent $i$-band magnitude. 3) Spectral slope calculated using the $g$ and $i$ measurements normalized to V-band following \cite{2012A&A...546A.115H}.
    \end{tablenotes}
\end{table*}

In the course of our observing campaign, we obtained optical $g/i$ photometry of 34 objects, 24 of which were $q<25$ AU KBOs, with the P200 telescope at Palomar Observatory and the Keck I 10 meter telescope at Maunakea Observatory. Observations taken at Palomar observatory used the WaSP instrument (Wafer-Scale Imager for Prime) and were collected in three days spread over 2018-20. Observations taken at Maunakea observatory used the LRIS instrument (Low Resolution Imaging Spectrometer) and were collected in two days, one in 2021 and one in 2022. Observations were done in series pairs of $g,i$ or $i,g$ observations from which colors were derived, with at least two but typically four colors obtained for each object, with two sets done an hour apart to confirm object on-sky motion. 

Reductions were performed using standard bias subtraction and flat fielding, followed by standard aperture photometry done using Pan-STARRS stars in the field of view of the images to calculate zero points. Apertures were selected to maximize the signal-to-noise, typically close to the FWHM seeing (typical values around 1.5'' at Palomar and 1.0'' at Keck). In our images, the average trailing of objects was sub-pixel for the Palomar observations and around one pixel for the Keck observations, with few objects trailing more than 2 pixels. Background photometric stars were characterized and selected using the \texttt{DAOfind} algorithm \citep{1987PASP...99..191S} as implemented in the \texttt{Photutils} package \citep{larry_bradley_2023_7946442}. Stars fainter than m$_g = 21$, with errors more than 0.03 mags, or with fewer than 5 observations per band were excluded from the zero-point calculations. Due to the differences between SDSS and Pan-STARRS filters, magnitudes were converted using the bandpass transformations given in \cite{2012ApJ...750...99T}. Background subtraction was performed with \texttt{SExtractor} \citep{1996A&AS..117..393B}. A notable point in the reduction is that we only use one of the four available detectors in the P200 WaSP images, as bias drift over the night causes each detector to scale in a visibly different way even after biasing, flat fielding, and background subtraction. We summarize the results of our observations in \autoref{tab:obspal} and \autoref{tab:obskeck}. The errors provided for the observations are obtained by calculating the standard deviation from multiple measurements of the color of the object, rather than from poisson noise on the photometry, as the photometric error for a set of measurements in any one band is dwarfed by the variability of an object and observing conditions over the course of a night. We convert our $g-i$ colors to spectral slopes normalized to $V$ band at 544.8 nm, following \cite{2012A&A...546A.115H}. Solar colors for $g$ and $i$ band are taken from \cite{2018ApJS..236...47W}.

\bibliography{ref}{}
\bibliographystyle{aasjournal}

\end{document}